\documentclass[11pt]{article}
\usepackage{mathrsfs}
\usepackage{graphicx}
\usepackage{CJK}
\begin{document}
\begin{CJK*}{GBK}{song}

\title{Excitation Spectrum and Momentum Distribution of Bose-Hubbard Model with On-site Two- and Three-body Interaction }
\author{Beibing Huang\\
Department of Experiment Teaching, Yancheng Institute of Technology,
Yancheng, 224051, China
\\ and Shaolong Wan \\
Institute for Theoretical Physics and Department of Modern Physics, University of Science and\\
Technology of China, Hefei, 230026, China}

\maketitle
\begin{center}
\begin{minipage}{120mm}
\vskip 0.8in
\begin{center}{\bf Abstract} \end{center}

{An effective action for Bose-Hubbard model with two- and three-body
on-site interaction in a square optical lattice is derived in the
frame of a strong-coupling approach developed by Sengupta and
Dupuis. From this effective action, superfluid-Mott insulator (MI)
phase transition, excitation spectrum and momentum distribution for
two phases are calculated by taking into account Gaussian
fluctuation about the saddle-point approximation. In addition the
effects of three-body interaction are also discussed.}

\end{minipage}
\end{center}

\vskip 1cm

\textbf{PACS} number(s): 03.75.Hh, 73.43.Nq, 05.30.Jp
\\

\section{Introduction}

The quantum phase transition from superfluid to Mott insulator (MI)
has been realized for cold bosonic atoms in a cubic optical lattice
\cite{greiner} and well explained in the frame of Bose-Hubbard model
(BHM) only with on-site two-body interaction \cite{fisher}: when the
lattice potential is very weak, the kinetic energy dominates and the
system is a delocalized superfluid. By contrast if the lattice
potential is very strong, the interaction will forbid bosons to
hopping from one site to another site and the system is in the MI
phase. Hence for intermediate lattice potential we naturally expect
the phase transition from superfluid to MI phase. In fact many other
phenomena of bosons in optical lattices can also be well described
by BHM \cite{jaksch1, jaksch2}.

However, the work of K\"{o}hler \cite{kohler} and Johnson {\it et
al.} \cite{johnson} pointed out that an effective three-body
interaction should be included to explain the experimental criterion
for collapse and revival of coherent matter waves \cite{mgreiner,
jpb1, jpb2}. In terms of a pure bosonic system, the effective
three-body interaction $g_3$ is related to the two-body scattering
length $a_s$ and the dilute gas parameter $\eta=\sqrt{\rho a_s^3}$
by $g_3\propto a_s^4\ln{[C\eta^2]}$ \cite{kohler, wu, braaten, fkh},
where $\rho$ is the boson density and the constant $C$ has been
determined by applying a microscopic description \cite{kohler}.
Generally in the current experiments the strength of three-body
interaction is much weaker than its two-body counterpart. In order
to enhance the ratio of three-body to two-body interactions, a
feasible route is by the induction of fermions. Considering the
mixture of a BEC and a single component Fermi gas, the Fermi degrees
of freedom can be completely integrated out and the effective
three-body interaction can be produced \cite{stchui, timmermans}. In
this method the interspecies interaction $g_{BF}$ not only induces
three-body interaction $g_3\propto g_{BF}^3$, but also weakens
two-body interaction to $g_2=g_B- P g_{BF}^2$, where $g_B$ and $g_2$
are bosonic two-body interaction before and after the integration of
Fermi degrees of freedom and $P$ is a constant \cite{stchui}. Very
explicitly by Feshbach resonance to tune interspecies interaction
$g_{BF}$ \cite{ferlaino, stan}, one should have enough space to
adjust the $g_3/g_2$. From this viewpoint it is very significant to
include three-body interaction into BHM.

As a matter of fact, some references \cite{chen, huang, kezhao,
india} have studied superfluid-MI phase transition in the presence
of two- and three-body on-site interactions at the mean field level
and found some interesting phenomena such as the rotation of phase
boundary around a fixed point and extension of MI. Especially the
reference \cite{kezhao} also studied the excitation spectrum in MI
by functional-integral method. Different from these works, in this
paper, we discuss the effects of three-body interaction on
excitation spectrum and momentum distribution both in the superfluid
and MI phases on the same footing, following a strong-coupling
approach \cite{sengupta, dupuis}. From the experimental viewpoint,
the momentum distribution can be measured by imaging atom gas after
turning off the trap potential, while the excitation spectrum
resulting from a particle-hole excitation can be attained by
applying a potential gradient to the system in the MI phase. It is
by probing the momentum distribution and excitation spectrum that
superfluid-MI phase transition is observed \cite{greiner}. In
section 2 we derive an effective action by performing two successive
Hubbard-Stratonovich transformations of the intersite hopping term.
In section 3 starting from this effective action, the phase diagram,
excitation spectrum and momentum distribution are calculated by
taking into account Gaussian fluctuations about the mean-field
approximation as in the Bogoliubov theories of the weakly
interacting Bose gas. A brief conclusion is given in section 4.
Throughout this paper, we set $\hbar = k_B = 1$, and take the
lattice constant unity.

\section{Hamiltonian and the Effective Action}

The system of bosons in an optical lattice with on-site two- and
three-body interaction can be described by the modified Bose-Hubbard
model
\begin{eqnarray}
H=-\sum_{i,j}t_{ij}a_i^{\dag}a_j + \frac{U}{2}\sum_i n_i(n_i-1) +
\frac{W}{6}\sum_i n_i (n_i -1)(n_i-2)-\mu \sum_i a_i^{\dag}a_i,
\label{1}
\end{eqnarray}
where $a_i$ is the bosonic field operator, $n_i=a_i^{\dag}a_i$ is
the particle number operator for bosons at the lattice site $i$. The
intersite hopping matrix $t_{ij}$ is nonzero (labeled $t$) only when
lattice sites $i$ and $j$ are the nearest neighbors, $U$, $W$ are
two- and three-body repulsive interaction strength among bosons. In
the light of the statement in the introduction about the strengths
of two- and three-body interactions, one consider that the ratio
$W/U$ is adjustable in a large parameter space. The last term
involving the chemical potential $\mu$ is added because it is very
convenient in the grand canonical ensemble. Without loss of
generality a square optical lattice is assumed.

In the imaginary-time functional integral formalism, the partition
function of the system is written as \cite{negele}
\begin{eqnarray}
Z&=&\int \mathscr{D}[a^{\ast}, a]e^{-S_0-S'},\nonumber\\
S_0&=&\int_0^{\beta}d\tau \sum_i a_i^{\ast}(\partial_{\tau}-\mu)a_i+
\frac{U}{2}\sum_i n_i(n_i-1) + \frac{W}{6}\sum_i n_i (n_i
-1)(n_i-2),\nonumber\\
S'&=&\int_0^{\beta}d\tau \left[-t\sum_{<i,j>}a_i^{\dag}a_j
\right],\label{2}
\end{eqnarray}
with $\beta=1/T$\ and $\tau$\ imaginary time. $\mathscr{D}[\cdots]$
represents the functional integral for field operators.

Carrying out Hubbard-Stratonovich transformation for $S'$ and
integrating out $a$, the partition function
\begin{eqnarray}
Z&=&\int\mathscr{D}[a^{\ast}, a, b^{\ast},
b]e^{-S_0-b_i^{\ast}t^{-1}_{ij}b_j + a^{\ast}_i b_i+ b^{\ast}_i
a_i},\nonumber\\
&=&Z_0\int\mathscr{D}[b^{\ast}, b]e^{-b_i^{\ast}t^{-1}_{ij}b_j}
<\sum_{n=0}^{\infty}\frac{(a^{\ast}_i b_i+ b^{\ast}_i
a_i)^n}{n!}>_0\nonumber\\
&=&Z_0\int\mathscr{D}[b^{\ast}, b]e^{-b_i^{\ast}t^{-1}_{ij}b_j +
V[b^{\ast}, b]}\label{3}
\end{eqnarray}
where $t^{-1}$ denotes the inverse matrix of the intersite hopping
matrix $t_{ij}$, $Z_0=\int\mathscr{D}[a^{\ast}, a]e^{-S_0}$ is the
partition function in the local limit ($t=0$), $<\cdots>_0$ means
that the average is taken with $S_0$. According to the linked
cluster expansion \cite{negele}, the functional $V[b^{\ast}, b]$
only includes contributions from linked diagrams. Until the fourth
order of $b$ and after a careful calculation
\begin{eqnarray}
V[b^{\ast}, b]&=&-\int_0^{\beta}d\tau_1
d\tau_2\sum_{i}b_i^{\ast}(\tau_1)G_{1c}(\tau_1,\tau_2)
b_i(\tau_2)+\nonumber\\&&\frac{1}{4}\int_0^{\beta}d\tau_1 d\tau_2
d\tau_3 d\tau_4 \sum_{i}
G_{2c}(\tau_1,\tau_2,\tau_3,\tau_4)b_i^{\ast}(\tau_1)
b_i^{\ast}(\tau_2) b_i(\tau_4) b_i(\tau_3),\label{4}
\end{eqnarray}
where $G_{1c}(\tau_1,\tau_2)$
($G_{2c}(\tau_1,\tau_2,\tau_3,\tau_4)$) denote the connected local
single (two) particle(s) Green function. In fact, $V[b^{\ast}, b]$
is the generating functional of connected Green functions in the
local limit if $b$ field is regarded as external source.

It is easily found that the relation between the Green function of
$a$ field $G_a$ and $G_b$ of $b$ field is
$G_a=t^{-1}-t^{-1}G_bt^{-1}$. Hence once $G_b$ is known, we can get
$G_a$. Unfortunately Green function $G_a$ obtained in this way is
not physical since it leads in the superfluid phase to a spectral
function which is not normalized to unity \cite{sengupta}. To
overcome this difficulty, a second Hubbard-Stratonovich
transformation was performed to decouple intersite hopping term and
the partition function becomes
\begin{eqnarray}
Z&=&\int\mathscr{D}[c^{\ast}, c, b^{\ast}, b]e^{c_i^{\ast}t_{ij}c_j
- c^{\ast}_i b_i- b^{\ast}_i c_i+V[b^{\ast}, b]}.\label{5}
\end{eqnarray}
Reference \cite{sengupta} has shown that the auxiliary field $c$ has
the same correlation functions as the original boson field $a$,
hence below we will substitute $a$ for $c$. In (\ref{5}) considering
the second order term in $V[b^{\ast}, b]$ as weight and integrating
out $b$ field, till the fourth order of $a$, the effective action
$S[a^{\ast}, a]$ is
\begin{eqnarray}
S[a^{\ast}, a]&=&-\int_0^{\beta}d\tau_1 d\tau_2\sum_{i,j}
a_i^{\ast}(\tau_1)\left[G_{1c}^{-1}(\tau_1,\tau_2)+t_{ij}\delta(\tau_1-\tau_2)\right]a_j(\tau_2)
\nonumber\\
&&+\frac{1}{4}\int_0^{\beta}d\tau_1 d\tau_2 d\tau_3 d\tau_4 \sum_{i}
\Gamma(\tau_1,\tau_2,\tau_3,\tau_4)a_i^{\ast}(\tau_1)
a_i^{\ast}(\tau_2) a_i(\tau_4) a_i(\tau_3),\label{6}
\end{eqnarray}
where we have discarded the contributions from all anomalous terms
in the same spirit of Ref.\cite{dupuis}.
$\Gamma(\tau_1,\tau_2,\tau_3,\tau_4)$ is the exact two-particle
vertex in the local limit
\begin{eqnarray}
\Gamma(\tau_1,\tau_2,\tau_3,\tau_4)&=& -\int_0^{\beta}d\tau_1'
d\tau_2' d\tau_3' d\tau_4'
G_{2c}(\tau_1',\tau_2',\tau_3',\tau_4')\times
\nonumber\\&&G_{1c}^{-1}(\tau_1,\tau_1')G_{1c}^{-1}(\tau_2,\tau_2')G_{1c}^{-1}(\tau_4,\tau_4')
G_{1c}^{-1}(\tau_3,\tau_3'),\label{7}
\end{eqnarray}
and has a complicated time-dependence. Taking the static limit of
$\Gamma$ into consideration and introducing a parameter
\begin{eqnarray}
g=\frac{1}{2}\Gamma_{static}=-\frac{1}{2}\frac{\overline{G}_{2c}}{\overline{G}_{1c}^4},\label{8}
\end{eqnarray}
where $\overline{G}_{1c}$, $\overline{G}_{2c}$ denote the
zero-frequency component of Fourier transformation of $G_{1c}$,
$G_{2c}$ respectively, we finally obtain
\begin{eqnarray}
S[a^{\ast}, a]&=&-\int_0^{\beta}d\tau_1 d\tau_2\sum_{i,j}
a_i^{\ast}(\tau_1)\left[G_{1c}^{-1}(\tau_1,\tau_2)+t_{ij}\delta(\tau_1-\tau_2)\right]a_j(\tau_2)
\nonumber\\&&+\frac{g}{2}\int_0^{\beta}d\tau\sum_{i} a_i^{\ast}
a_i^{\ast} a_i a_i.\label{9}
\end{eqnarray}
In contrast to the original action (\ref{2}), on the one hand the
intersite hopping has been renormalized by the exact local single
particle Green function $G_{1c}$, on the other hand the interaction
$U$ has been substituted by the exact local two-particle vertex.
This action (\ref{9}) is the starting point of our analysis.

\section{Excitation Spectrum and Momentum Distribution}

In this section we decide the boundary of superfluid-MI phase
transition, excitation spectrum and momentum distribution for both
superfluid phase and MI at zero temperature. Before doing these, we
first calculate local Green functions $G_{1c}$ and $G_{2c}$ at zero
temperature to determine the parameter $g$.

In the local limit, the Hamiltonian (\ref{1}) is diagonal about
lattice sites and its eigenstates are Fock states
$|n>=(n!)^{-1/2}(a^{\dag})^n|0>$ with eigenvalue
$\epsilon_n=(U/2)n(n-1)+(W/6)n(n-1)(n-2)-\mu n$. In order to attain
$g$ we firstly calculate one-particle Green function
$G_1(\tau)=-<\mathcal {T}_{\tau}a(\tau)a^{\dag}(0)>$ and
two-particle Green function
$G_2(\tau_1,\tau_2;\tau_3,\tau_4=0)=<\mathcal
{T}_{\tau}a(\tau_1)a(\tau_2)a^{\dag}(0)a^{\dag}(\tau_3)>$, which are
easily calculated using the closure relation
$\sum_{n=0}^{\infty}|n><n|=1$. Then $G_{1c}(\tau)=G_1(\tau)$ and
$G_{2c}(\tau_1,\tau_2;\tau_3,\tau_4)=G_2(\tau_1,\tau_2;\tau_3,\tau_4)
-G_1(\tau_1,\tau_3)G_1(\tau_2,\tau_4)-G_1(\tau_1,\tau_4)G_1(\tau_2,\tau_3)$.
Making Fourier transformation, we obtain
\begin{eqnarray}
\overline{G}_{1c}(i\omega_n)&=&\frac{-n_0}{i\omega_n+\epsilon_{n_0-1}-\epsilon_{n_0}}+
\frac{n_0+1}{i\omega_n+\epsilon_{n_0}-\epsilon_{n_0+1}},\label{10}\\
\overline{G}_{2c}&=&-\frac{4(n_0+1)(n_0+2)}{(\epsilon_{n_0}-\epsilon_{n_0+2})
(\epsilon_{n_0}-\epsilon_{n_0+1})^2}-\frac{4n_0(n_0-1)}{(\epsilon_{n_0}-\epsilon_{n_0-2})
(\epsilon_{n_0}-\epsilon_{n_0-1})^2}\nonumber\\&+&\frac{4n_0(n_0+1)}{(\epsilon_{n_0}-\epsilon_{n_0+1})
(\epsilon_{n_0}-\epsilon_{n_0-1})^2}+\frac{4n_0(n_0+1)}{(\epsilon_{n_0}-\epsilon_{n_0+1})^2
(\epsilon_{n_0}-\epsilon_{n_0-1})}\nonumber\\&+&\frac{4n_0^2}{(\epsilon_{n_0}-\epsilon_{n_0-1})^3
}+\frac{4(n_0+1)^2}{(\epsilon_{n_0}-\epsilon_{n_0+1})^3},\label{11}
\end{eqnarray}
where $n_0$ is the occupation number of ground state in the local
limit which minimizes the eigenvalue $\epsilon_n$.

\subsection{Superfluid-MI Phase Transition}

To proceed further, we consider a quadratic expansion of the action
(\ref{9}) in terms of fluctuation near the saddle-point value of $a$
field. Choosing saddle-point value $\psi_0$ for $a$, that is
$a_i(\tau)\rightarrow \psi_0 + \widetilde{a}_i(\tau)$, (\ref{9})
becomes $S=S_0+S_1+S_2$ with
\begin{eqnarray}
S_0&=&-(\overline{G}_{1c}^{-1}+4t)\psi_0^2+
\frac{g}{2}\psi_0^4,\nonumber\\
S_1&=&-\psi_0(\overline{G}_{1c}^{-1}+4t-g\psi_0^2)\int_0^{\beta}d\tau\sum_i
[\widetilde{a}_i(\tau)+\widetilde{a}_i^{\ast}(\tau)],\nonumber\\
S_2&=&-\int_0^{\beta}d\tau_1 d\tau_2\sum_{i,j}
\widetilde{a}_i^{\ast}(\tau_1)\left[G_{1c}^{-1}(\tau_1,\tau_2)+t_{ij}\delta(\tau_1-\tau_2)\right]\widetilde{a}_j(\tau_2)
\nonumber\\&&+\frac{g}{2}\psi_0^2\int_0^{\beta}d\tau\sum_{i}
\widetilde{a}_i^{\ast} \widetilde{a}_i^{\ast} + \widetilde{a}_i
\widetilde{a}_i + 4 \widetilde{a}_i^{\ast}
\widetilde{a}_i.\label{12}
\end{eqnarray}
The constant $\psi_0$ is determined by requiring the coefficients of
the linear terms in $\widetilde{a}_i$ and $\widetilde{a}_i^{\ast}$
to vanish, leading to the equation
\begin{eqnarray}
\overline{G}_{1c}^{-1}+4t-g\psi_0^2=0.\label{13}
\end{eqnarray}
It is well-known that a nonzero $\psi_0$ signifies superfluid of
system, therefore the boundary of superfluid-MI phase transition is
\begin{eqnarray}
\overline{G}_{1c}^{-1}+4t=0,\label{14}
\end{eqnarray}
which is in agreement with the results in \cite{chen, huang, kezhao,
india}. Fig. 1, the phase diagram in the $t/U-\mu/U$ plane for
different $W/U$ from (\ref{14}), shows that the effects of
three-body interaction on the phase boundary are very dramatic. On
the one hand Mott lobes with $n_0= 1$ remains unaltered in the
presence of $W$. On the other hand for Mott lobes with higher
densities, with the increase of $W$, their widths and heights
gradually increase. In addition for small $W$ phase boundary has the
same characteristics as the conventional superfluid- MI phase
transition that Mott lobes with higher occupation have smaller area.
While $W$ gradually increases and dominates two-body interaction
$U$, Mott lobes with higher occupation have bigger area. These
phenomena can be illustrated from the pure two-body and three-body
interaction \cite{chen}.

\begin{figure}[tbp]
\centering
\includegraphics[width=7.5cm, height=6.0cm]{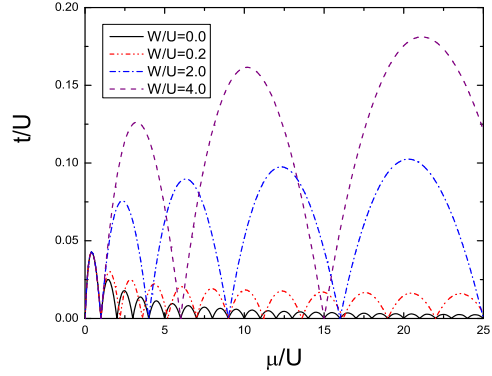}
\caption{The phase diagram of superfluid-MI phase transition with
different three-body interaction $W$. } \label{fig.1}
\end{figure}

The excitation spectrum and momentum distribution can be deduced
from the second order action $S_2$. Introducing the Fourier
transformation
\begin{eqnarray}
\widetilde{a}(\vec{k}, i\omega_n)=\frac{1}{\sqrt{\beta
N}}\int_0^{\beta}d\tau\sum_i\widetilde{a}_i(\tau)
e^{-i\vec{k}\cdot\vec{R}_i+i\omega_n\tau},\label{15}
\end{eqnarray}
where $N$ is the total number of lattice sites, $\vec{R}_i$ is a
lattice vector for site $i$, $\omega_n=2n\pi/\beta$ is bosonic
Matsubara frequency, $S_2$ becomes
\begin{eqnarray}
S_2&=&\frac{1}{2}\sum_{\vec{k},\omega_n}\left[\widetilde{a}^{\ast}(\vec{k},
i\omega_n),\widetilde{a}(-\vec{k},
-i\omega_n)\right]M_{2\times2}\left[\widetilde{a}(\vec{k},
i\omega_n),\widetilde{a}^{\ast}(-\vec{k}, -i\omega_n)\right]
\label{16}
\end{eqnarray}
with
\begin{eqnarray}
M_{2\times2}&=&\left(
\begin{array}{cc}
-G_{1c}^{-1}(i\omega_n)+\varepsilon_k+2g\psi_0^2 & g\psi_0^2 \\
\\
g\psi_0^2 & -G_{1c}^{-1}(-i\omega_n)+\varepsilon_{-k}+2g\psi_0^2
\end{array} \right)\label{17}
\end{eqnarray}
and $\varepsilon_{k}=-2t(\cos{k_x}+\cos{k_y})$ denoting the boson
dispersion in the absence of on-site interaction.

\subsection{MI Phase}

\begin{figure}[tbp]
\includegraphics[width=7.5cm, height=6.0cm]{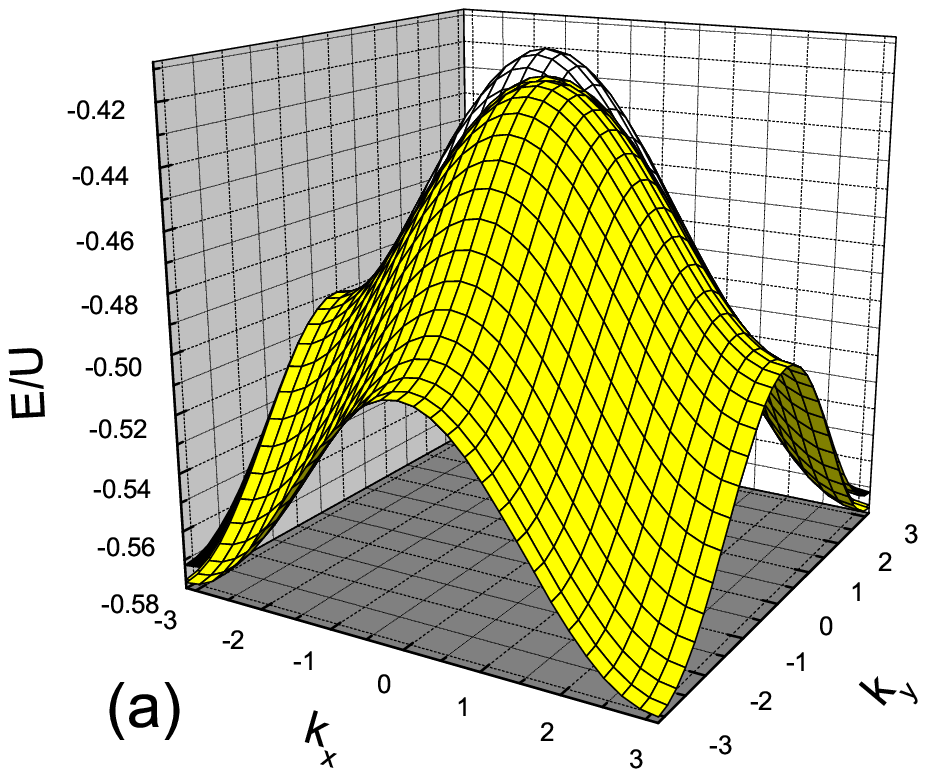}
\includegraphics[width=7.5cm, height=6.0cm]{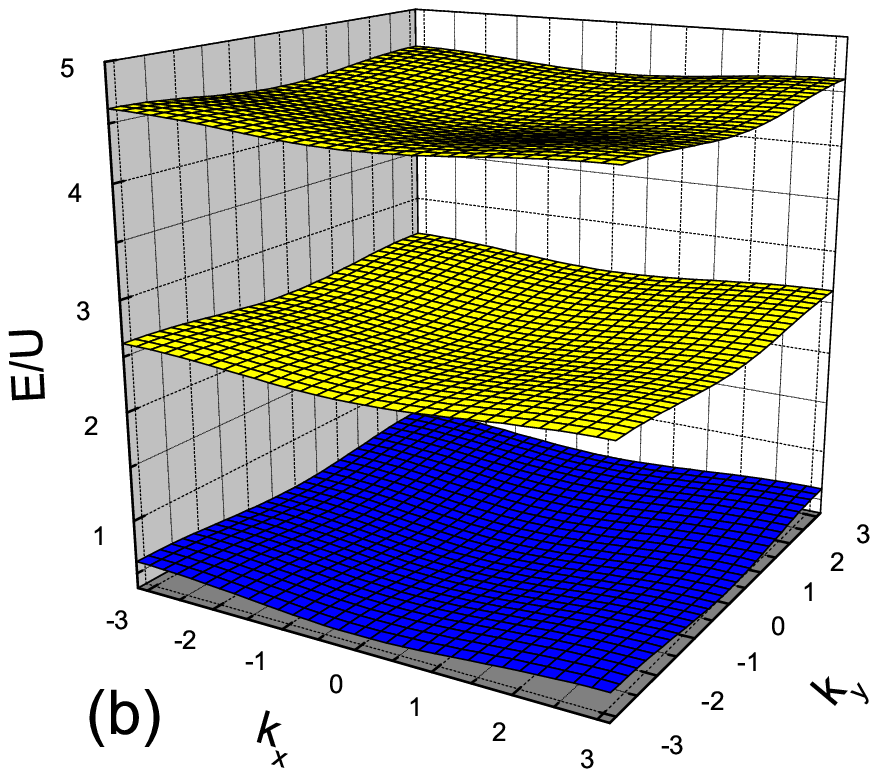}
\caption{The excitation spectrum for quasiholes (a) and
quasiparticles (b). We choose $t/U=0.01$, $\mu/U=1.5$ and $W/U=0,
2.0, 4.0$ to make the system into MI with the filling factor
$n_0=2$. In (a) the effects of different $W$ is very small. In (b)
from bottom to top $W$ is increasing. } \label{fig.2}
\end{figure}

In the MI phase $\psi_0=0$, the matrix $M_{2\times2}$ becomes
diagonal and $S_2$ is reduced into
\begin{eqnarray}
S_2=\sum_{\vec{k},\omega_n}\widetilde{a}^{\ast}(\vec{k},
i\omega_n)\left[-G_{1c}^{-1}(i\omega_n)+\varepsilon_{k}\right]\widetilde{a}(\vec{k},
i\omega_n).\label{18}
\end{eqnarray}
The Green function $g_{MI}(\vec{k},
i\omega_n)=-<\widetilde{a}(\vec{k},
i\omega_n)\widetilde{a}^{\ast}(\vec{k}, i\omega_n)>$ can be directly
obtained $g_{MI}^{-1}(\vec{k},
i\omega_n)=G_{1c}^{-1}(i\omega_n)-\varepsilon_{{k}}$. Using
(\ref{10}), we obtain
\begin{eqnarray}
g_{MI}(\vec{k},
i\omega_n)=\frac{z_k}{i\omega_n-E_k^+}+\frac{1-z_k}{i\omega_n-E_k^-}\label{19}
\end{eqnarray}
with
\begin{eqnarray}
E_k^{\pm}&=&\frac{\epsilon_{n_0+1}-\epsilon_{n_0-1}+\varepsilon_k\pm\sqrt{\Delta_k}}{2},\label{20}\\
\Delta_k&=&(\epsilon_{n_0+1}-\epsilon_{n_0-1}+\varepsilon_k)^2-4(\epsilon_{n_0-1}-\epsilon_{n_0})
(\epsilon_{n_0}-\epsilon_{n_0+1})\nonumber\\&-&4\varepsilon_k[n_0(\epsilon_{n_0}-\epsilon_{n_0+1})+
(n_0+1)(\epsilon_{n_0}-\epsilon_{n_0-1})],\label{21}\\
z_k&=&\frac{E_k^+ - n_0(\epsilon_{n_0}-\epsilon_{n_0+1})-(n_0+1)
(\epsilon_{n_0}-\epsilon_{n_0-1})}{E_k^+-E_k^-}.\label{22}
\end{eqnarray}

In (\ref{20}), $E_{k}^{\mp}$ stand for excitations of quasiholes
(removing particles from sites) or  quasiparticles (adding particles
to sites), which is consistent with the results in \cite{chen} and
is shown in Fig.2(a) (Fig.2(b)). Very explicitly, three-body
interaction $W$ has little impacts on quasihole excitations, but it
has the large effects on quasiparticle excitation: with the increase
of $W$, quasiparticle energy becomes higher, hence its excitation
also becomes difficult.

\begin{figure}[tbp]
\centering
\includegraphics[width=7.5cm, height=6.0cm]{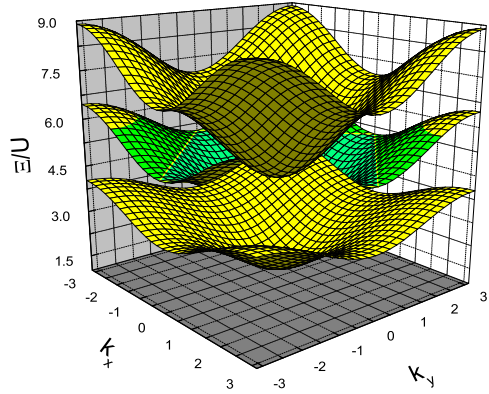}
\caption{The excitation spectrum for density fluctuation $\Xi_k$. We
choose $t/U=0.06$, $W/U=2.0$ and $\mu/U=1.5, 5.5, 10.5$ to make the
system into MI with filling factor $n_0=2,3,4$ respectively. From
bottom to top $\mu$ is increasing.} \label{fig.3}
\end{figure}

In Greiner's experiment \cite{greiner}, Superfluid-MI phase
transition is also detected by applying a potential gradient to the
system in the MI phase and probing the excitation spectrum resulting
from a particle-hole excitation. According to the above results, we
can find a first approximation for the dispersion of particle-hole
excitation (density fluctuations) $\Xi_k$ by subtracting $E_{k}^{-}$
from $E_{k}^{+}$, which yields
\begin{eqnarray}
\Xi_k=E_{k}^{+}-E_{k}^{-}=\sqrt{\Delta_k}. \label{23}
\end{eqnarray}
Due to little impacts of three-body interaction $W$ on $E_{k}^{-}$,
the dependence of $\Xi_k$ on $W$ is the same as $E_{k}^{+}$. In
Fig.3 we show $\Xi_k$ for a fixed $W$, but different filling factor
$n_0$, i.e. different chemical potential $\mu$. We can see that in
MI phases there is always a band gap increasing with the filling
factor, which is different from the situation without three-body
interaction where the gap is approximately independent of the
filling factor. It is from this perspective that reference
\cite{india} proposed to observe weak three-body interaction for
high filling MI.

\begin{figure}[tbp]
\centering
\includegraphics[width=7.5cm, height=6.0cm]{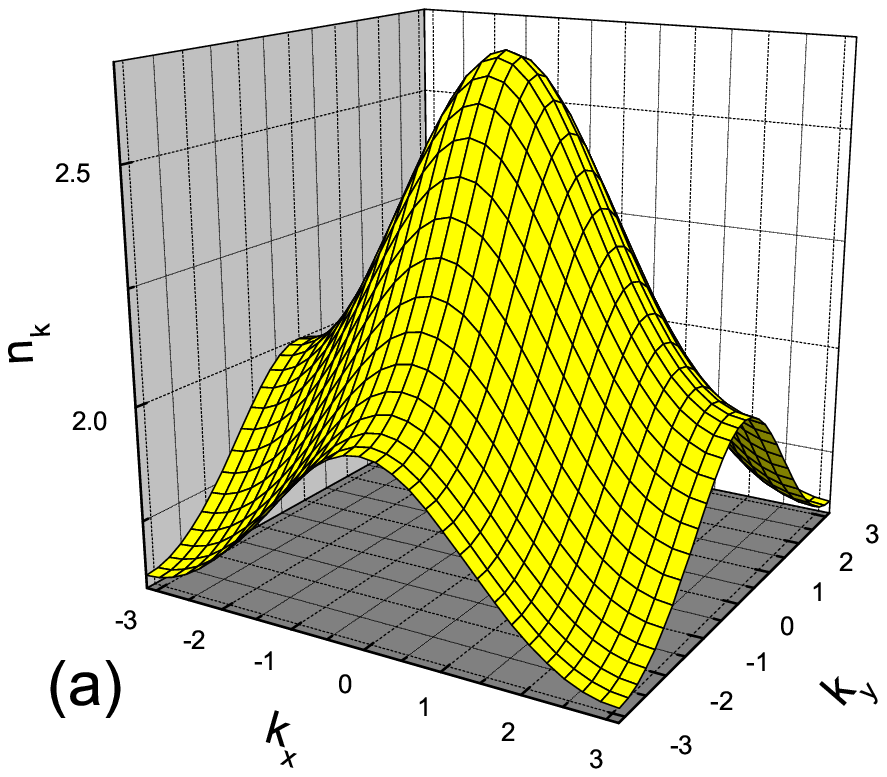}
\includegraphics[width=7.5cm, height=6.0cm]{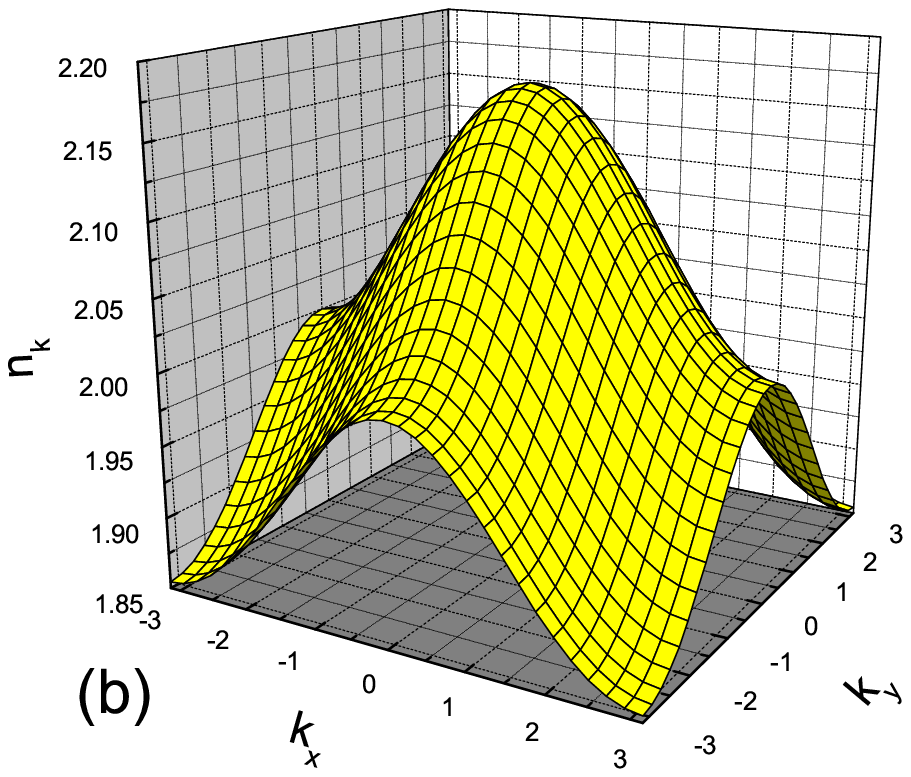}
\caption{The momentum distribution in MI phases with filling factor
$n_0=2.0$. Parameters are chosen as follows: $t/U=0.01$, $\mu/U=1.5$
and $W/U=0$ in (a) $W/U=2.0$ in (b).} \label{fig.4}
\end{figure}

From the theorem of fluctuation and dissipation, the momentum
distribution at zero temperature
$n_k=<\widetilde{a}^{\ast}(\vec{k})\widetilde{a}(\vec{k})>$ is
\begin{eqnarray}
n_k&=&\int_{-\infty}^{\infty}\frac{d\omega}{2\pi}\frac{1}{e^{\beta\omega}-1}
(-2)Im[g_{MI}(\vec{k},
\omega+i0^+)]\nonumber\\&=&\int_{-\infty}^{\infty}\frac{d\omega}{e^{\beta\omega}-1}[z_k\delta(\omega-E_k^+)
+(1-z_k)\delta(\omega-E_k^-)]=z_k-1, \label{24}
\end{eqnarray}
where the label "Im" denotes the imaginary part. In deriving
(\ref{24}) we have noted that quasiparticle dispersion $E_k^+$ is
always greater than or equal to zero and $E_k^-$ is always smaller
than or equal to zero. Because of this only the quasiholes give a
contribution to the total density at zero temperature. Fig.4 shows
momentum distributions in MI phases and we find that three-body
interaction $W$ leads to not only flattening but also broadening of
momentum distribution around $\vec{k}=0$.

\subsection{Superfluid Phase}

In terms of the superfluid phase, $\psi_0 ^2=
(\overline{G}_{1c}^{-1}+4t)/g$. From the condition $det(M_{2\times
2})=0$, we obtain four excitation branches $\pm E_k^{\pm}$ with
\begin{eqnarray}
E_k^{\pm
2}&=&-\frac{B_k}{2}\pm\frac{1}{2}(B_k^2-4C_k)^{1/2},\nonumber\\
B_k&=&2h_2-2(\varepsilon_k+2g\psi_0^2)h_1-(h_3-\varepsilon_k-2g\psi_0^2)^2
+g^2\psi_0^4,\nonumber\\
C_k&=&[h_2-(\varepsilon_k+2g\psi_0^2)h_1]^2-g^2\psi_0^4h_1^2,\nonumber\\
h_1&=&(n_0+1)(\epsilon_{n_0-1}-\epsilon_{n_0})-n_0(\epsilon_{n_0}-\epsilon_{n_0+1}),\nonumber\\
h_2&=&(\epsilon_{n_0-1}-\epsilon_{n_0})(\epsilon_{n_0}-\epsilon_{n_0+1}),\nonumber\\
h_3&=&(\epsilon_{n_0-1}-\epsilon_{n_0})+(\epsilon_{n_0}-\epsilon_{n_0+1}).\label{25}
\end{eqnarray}
Compared to the Bogoliubov's theory for weakly interacting Bose
gases, where only two excitation branches exist, there are two more
excitation branches $\pm E_k^+$ due to renormalization of the
tunneling term by locally exact single-particle Green function
$\overline{G}_{1c}(i\omega)$. Fig.5 show typical characters for
$E_k^{\pm}$: $E_k^+$ is gapped while $E_k^-$ is gapless. Moreover we
also find that a gap opens between $E_k^+$ and $E_k^-$ when
three-body interaction $W$ becomes strong. Here, we are only
interested in gapless excitations described by $\pm E_k^-$, which
originate from spontaneously broken gauge symmetry and describe the
low energy behaviors of the system. By expanding $E_k^-$ in the
vicinity of $\vec{k}=0$, we find a linear spectrum $E_k^-=vk$ with
\begin{eqnarray}
v=\left[\frac{2t(\overline{G}_{1c}^{-1}+4t)}{\alpha^2+2\gamma(\overline{G}_{1c}^{-1}+4t)}\right]^{1/2},
\alpha =\frac{h_2-h_1h_3}{h_1^2}, \gamma
=\frac{h_1^2+h_2-h_1h_3}{h_1^3}. \label{26}
\end{eqnarray}
The existence of such linear spectrum is consistent with the
Landau's criterion of superfluidity \cite{landau}. Fig.6 shows the
dependence of sound velocity $v$ on three-body interaction $W$. It
is should be noted that the system is in the superfluid phase near
$n_0=2$ MI phase for parameters $\mu$ and $t$ chosen in Fig.6. For
different parameters $\mu$ and $t$, on the one hand $v$ generally is
not monotonic as the function of $W$, on the other hand the critical
values $W_c$, where $v=0$ or the system enters into the MI phase,
are very different. The second character can be illustrated from the
phase diagram Fig.1: with the increase of $W$, the left boundary of
MI lobe with $n_0=2$ changes very slowly, but the right boundary
changes very quickly.

\begin{figure}[tbp]
\centering
\includegraphics[width=7.5cm, height=6.0cm]{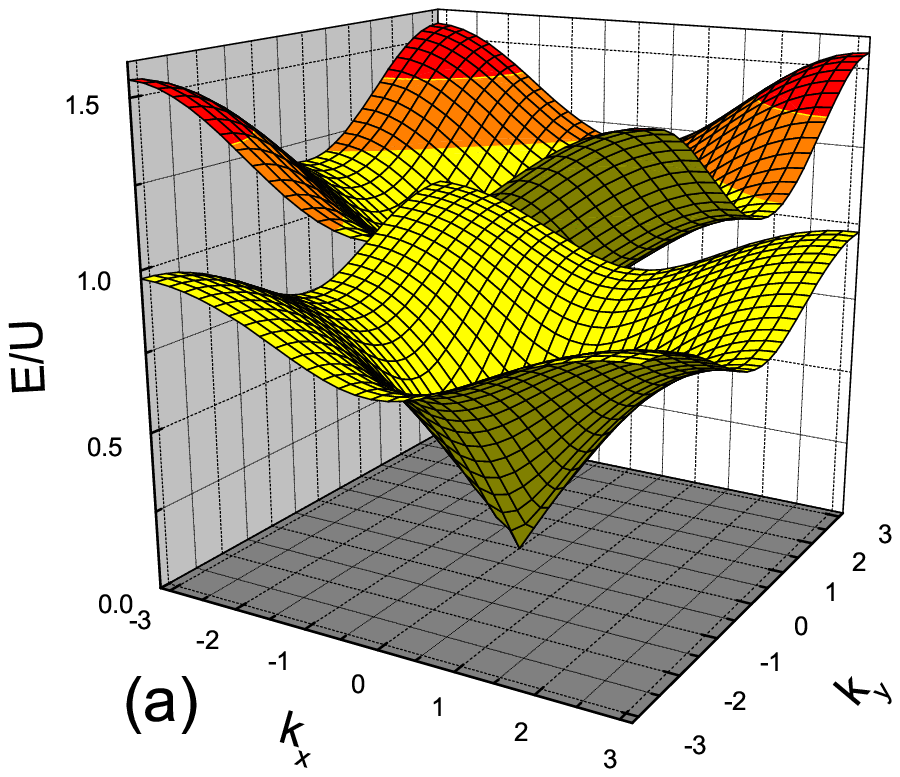}
\includegraphics[width=7.5cm, height=6.0cm]{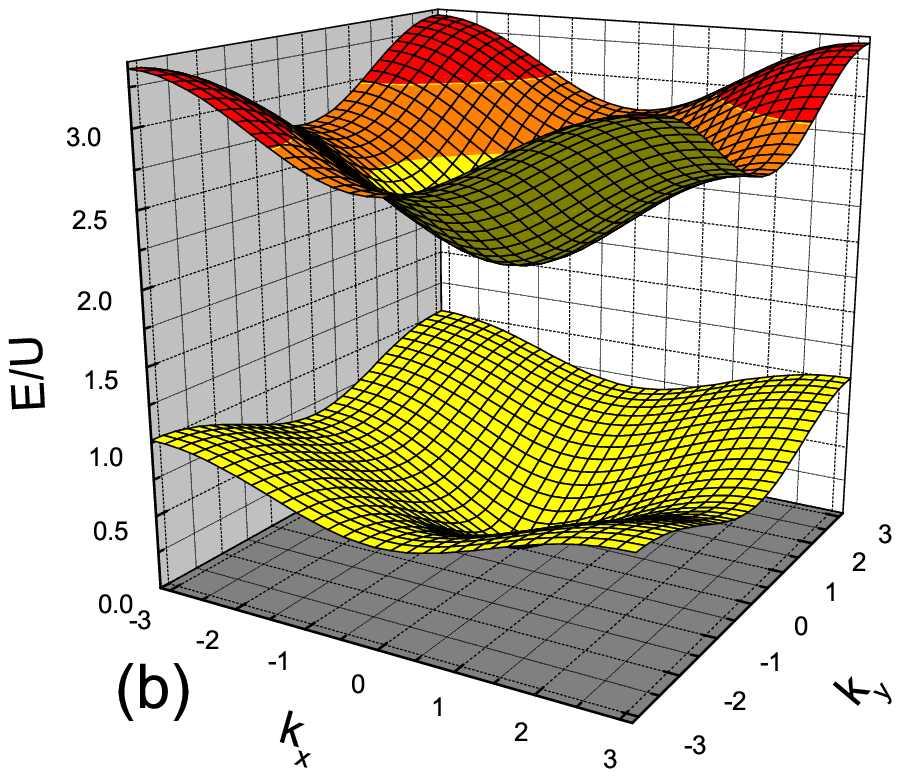}
\caption{The excitation spectrum in superfluid phase with parameters
$t/U=0.06$, $\mu/U=1.5$, $W/U=0.0$ in (a) and $W/U=2.0$ in (b).}
\label{fig.5}
\end{figure}

From the excitation spectrum $\pm E_k^{\pm}$, the Green function
$g_{SF}(\vec{k}, i\omega_n)$ can be easily obtained
\begin{eqnarray}
g_{SF}(\vec{k},
i\omega_n)=\frac{P_+}{i\omega_n+E_k^+}+\frac{P_-}{i\omega_n-E_k^+}
+\frac{Q_+}{i\omega_n+E_k^-}+\frac{Q_-}{i\omega_n-E_k^-},\label{27}
\end{eqnarray}
where
\begin{eqnarray}
P_{\pm}&=&\frac{E_k^{+3}\mp a_2E_k^{+2}-a_1E_k^{+}\mp
a_0}{2E_k^+(E_k^{+2}-E_k^{-2})}, Q_{\pm}=\frac{E_k^{-3}\mp
a_2E_k^{-2}-a_1E_k^{-}\mp
a_0}{2E_k^-(E_k^{-2}-E_k^{+2})},\nonumber\\
a_2&=&h_1-h_3+\varepsilon_k+2g\psi_0^2, a_1=h_3h_1-h_2,
a_0=h_1[h_2-h_1(\varepsilon_k+2g\psi_0^2)].\label{28}
\end{eqnarray}

\begin{figure}[tbp]
\centering
\includegraphics[width=5.0cm, height=4.5cm]{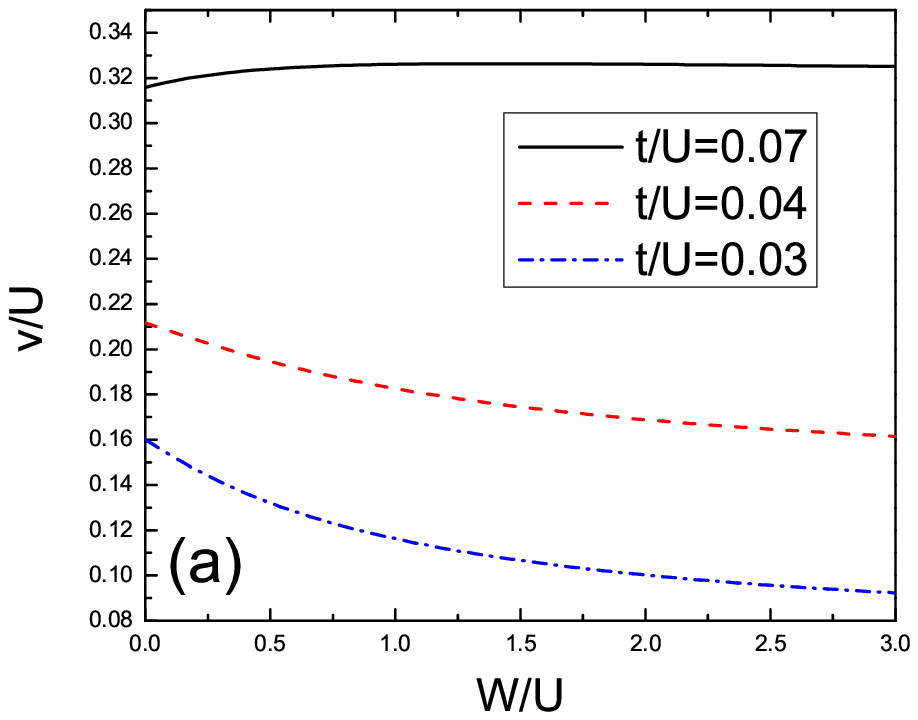}
\includegraphics[width=5.0cm, height=4.5cm]{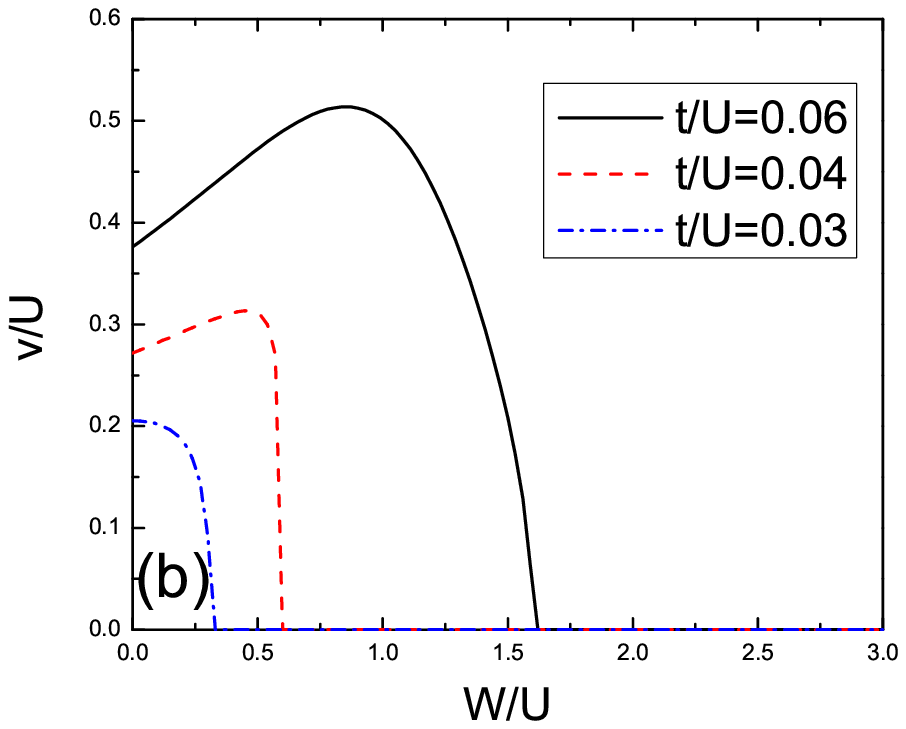}
\includegraphics[width=5.0cm, height=4.5cm]{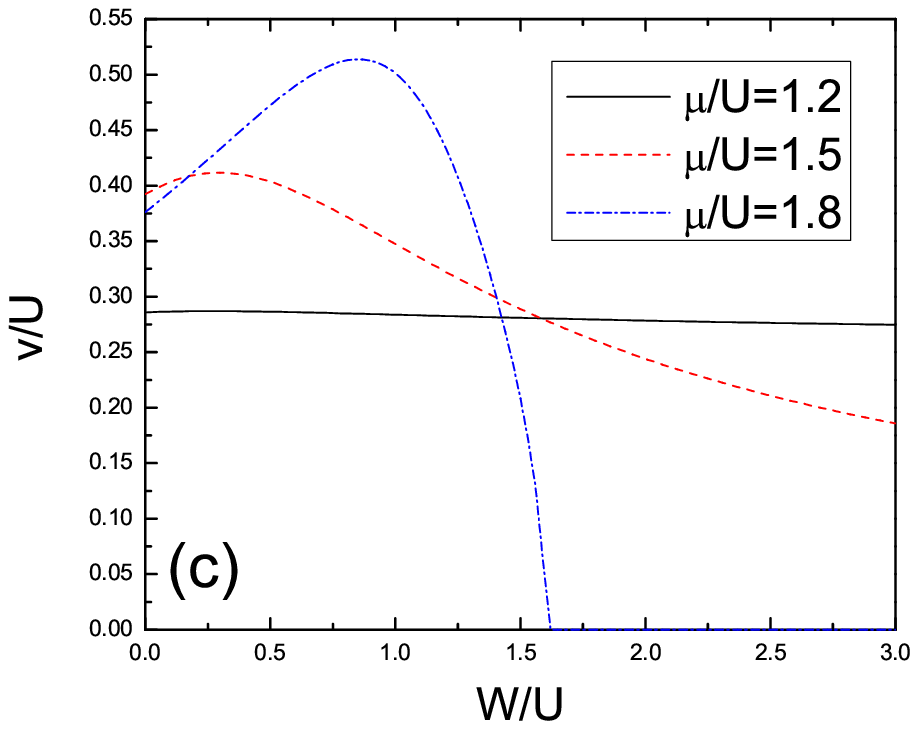}
\caption{The sound velocity $v$ as the function of three-body
interaction $W$ for (a) $\mu/U=1.2$, (b) $\mu/U=1.8$, (c)
$t/U=0.06$.} \label{fig.6}
\end{figure}

Following the same steps in (\ref{24}), the momentum distribution in
the superfluid phase is
\begin{eqnarray}
n_k=N\psi_0^2\delta_{k,0}-P_+-Q_+,\label{29}
\end{eqnarray}
where the first term $N\psi_0^2\delta_{k,0}$ comes from the
condensate and it is this term that supplies a coherent peak in
absorptive images of atom gases. Fig.7 show the momentum
distribution from the second and third terms. In contrast to the
situation in MI phases, three-body interaction $W$ only broadens the
width of the peak near $\vec{k}=0$.

\begin{figure}[tbp]
\centering
\includegraphics[width=7.5cm, height=6.0cm]{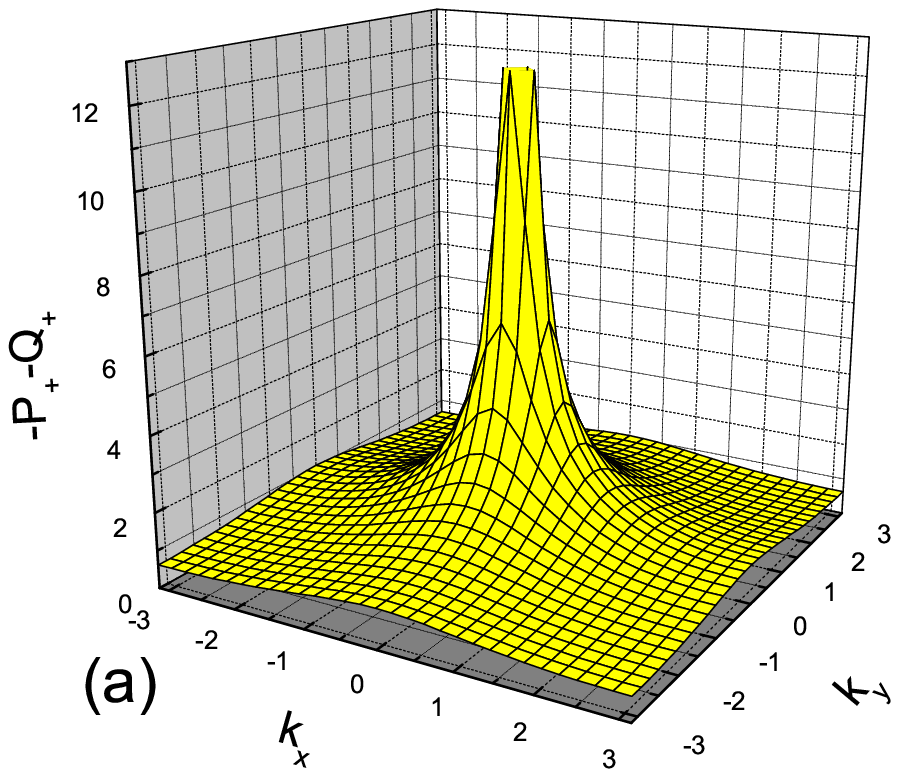}
\includegraphics[width=7.5cm, height=6.0cm]{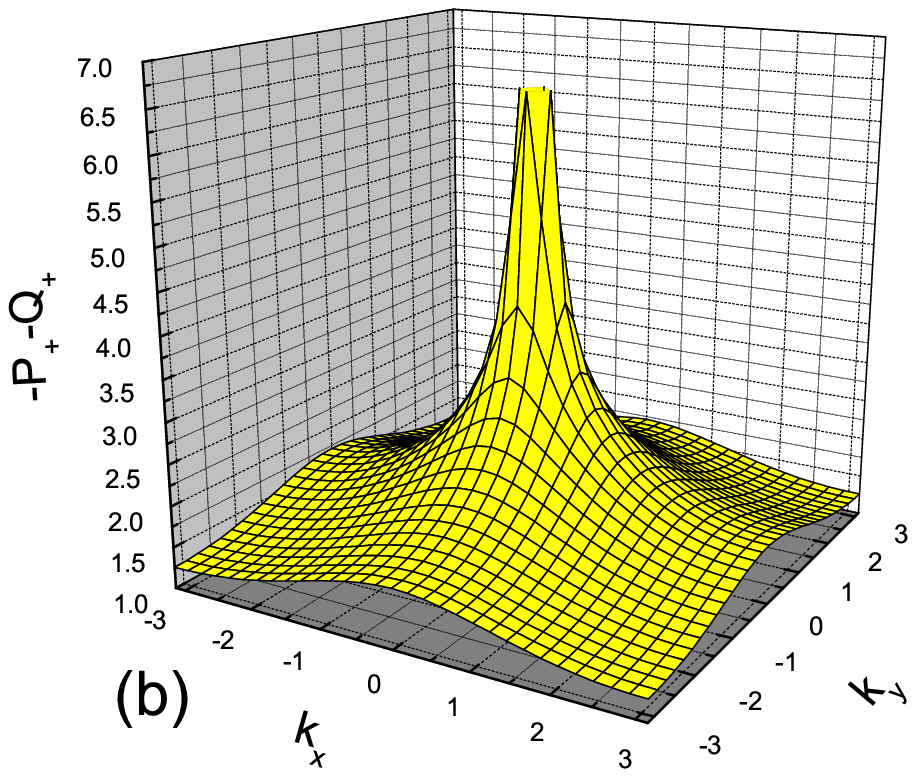}
\caption{The momentum distribution in superfluid phase. Parameters
are chosen as follows: $t/U=0.06$, $\mu/U=1.5$ and $W/U=0$ in (a)
$W/U=2.0$ in (b).} \label{fig.7}
\end{figure}

\section{Conclusions}

In conclusion, an effective action for Bose-Hubbard model with two-
and three-body on-site interaction in a square optical lattice has
been derived by performing two successive Hubbard-Stratonovich
transformations of the intersite hopping term. The main advantage of
this method is that superfluid and MI phases can be analyzed on the
same footing. Starting from this effective action, superfluid-MI
phase transition, excitation spectrum and momentum distribution for
two phases are calculated by taking into account Gaussian
fluctuation about the saddle-point approximation. In addition the
effects of three-body interaction are also discussed. We find that
the sound velocity in superfluid phase generally is not monotonic as
the function of three-body interaction, depending on the chemical
potential and hopping term.

\section*{Acknowledgement}

The work was supported by National Natural Science Foundation of
China under Grant No. 11275108. The author Huang also thanks
Foundation of Yancheng Institute of Technology under Grant No.
XKR2010007.

\end{CJK*}
\end{document}